\documentclass[%
 reprint,
 amsmath,amssymb,
 aps,
]{revtex4-2}

\usepackage{graphicx}
\usepackage{dcolumn}
\usepackage{bm}

\begin{document}

\preprint{APS/123-QED}

\title{Quantitative calibration of a TWPA applied to an optomechanical platform }

\author{Alexandre Delattre}

\author{Ilya Golokolenov}%

 \author{Richard Pedurand}

  \author{Nicolas Roch}

  \author{Arpit Ranadive$^\dag$}

   \author{Martina Esposito$^\ddag$}

   \author{Luca Planat$^\P$}

 \author{Andrew Fefferman}

  \author{Eddy Collin}
 \email{Eddy.Collin@neel.cnrs.fr}

\affiliation{Univ. Grenoble Alpes, Institut Néel - CNRS UPR2940, 25 rue des Martyrs, BP 166, 38042 Grenoble Cedex 9, France \\
$^\dag$ now at: Google QAI, Santa Barbara, CA, USA \\
$^\ddag$ now at: CNR-SPIN Complesso di Monte S. Angelo, via Cintia, Napoli 80126, Italy \\
$^\P$ now at: Silent Waves, 69-73 Rue Félix Esclangon, 38000 Grenoble, France}%

 \author{\vspace*{-0.2cm} Xin Zhou}
 \affiliation{IEMN, Univ. Lille - CNRS UMR8520, Av. Henri Poincaré, Villeneuve d’Ascq 59650, France}


 
\author{\vspace*{-0.2cm} Mika A. Sillanpää}
\author{Laure Mercier de Lépinay}

\affiliation{Department of Applied Physics, Aalto University, 00076, Espoo, Finland}%

\author{\vspace*{-0.2cm}  Andrew D. Armour}
\author{Jonas Glatthard}
\affiliation{School of Physics and Astronomy and 
Centre for the Mathematics and Theoretical Physics of Quantum Non-Equilibrium Systems, University of Nottingham, Nottingham NG7 2RD, UK}%

\date{\today}
\vspace*{0.2cm}

\begin{abstract}
In the last decade, the microwave quantum electronics toolbox has been enriched with quantum-limited detection devices such as Traveling Wave Parametric Amplifiers (TWPAs). 
The extreme sensitivity they provide is not only mandatory for some physics applications within quantum information processing, but is also the key element that will determine the detection limit of quantum sensing setups. 
In the framework of microwave optomechanical systems, an unprecedented range of small motions and forces is accessible, for which a  {\it specific quantitative} calibration becomes necessary.
We report on 
near quantum-limited measurements performed with an aluminum drumhead mechanical device within the temperature range $4~$mK $-$ $400~$mK. The whole setup is carefully calibrated, especially taking into account the power-dependence of microwave absorption in the superconducting optomechanical cavity. 
This effect is commonly attributed to Two-Level-Systems (TLSs) present in the metal oxide. We demonstrate that a similar feature exists in the TWPA, and can be phenomenologically fit with adapted expressions. 
If not taken into account, the error on the signal strength can be as large as a factor of about 2, which is unacceptable for quantitative experiments.
The power and temperature dependence is studied over the full parameter range, leading to {\it an absolute} definition 
 of phonon population  (i.e. Brownian motion amplitude), with an uncertainty $\pm 20~\%$
  limited by sources of noise internal to the optomechanical element. 

\begin{description}
\item[Keywords]
Quantum-limited amplifier, Microwave optomechanics, Two level systems
\end{description}
\end{abstract}

\maketitle


\section{\label{sec:level1}Introduction}

The advent of solid state quantum technologies has boosted cryogenic microwave engineering, and especially the detection capabilities.  Ultra-low noise High Electron Mobility Transistors (HEMT) are now commercially available \cite{lnf}, and 
near {\it quantum-limited} amplifiers have been developed.  
These, based on superconducting devices, originally included Josephson Bifurcation Amplifiers (JBA) and Josephson Parametric Amplifiers (JPA)  \cite{PhysRevB.76.014524,Vijay,PhysRevB.76.014525, Yurke,Roy}.
They reach the Standard Quantum limit (SQL) at Giga-Hertz frequencies, but suffer from a very narrow bandwidth 
unless {\it ad hoc} impedance engineering approaches are adopted \cite{mart1,mart2}.
To overcome this shortcoming, a new generation of quantum amplifiers has been developed: Traveling Wave Parametric Amplifiers (TWPA) \cite{PhysRevX.10.021021,Macklin,Farzad,White}, which are even now 
available as commercial products \cite{sw}. \\

TWPAs have become essential for obtaining 
the best possible fidelity in quantum bit readout \cite{PhysRevLett.112.170501, 2024arXiv240718778T}, a necessity if one is to build a functional quantum computer. 
But TWPAs are also a unique resource in quantum sensing: for instance, they are part of the circuitry used to build Single Microwave Photon Detectors (SMPD) \cite{Wang}, sensitive enough to read-out and control {\it single nuclear spins} \cite{Osullivan24}. 

For these usages, the TWPA detection enables the best possible discrimination between quantum bit states $|0\! >$ and $|1 \! >$. In this respect, the read-out is essentially relative. But specific quantum sensing issues, especially linked to fundamental research, require an {\it absolute quantitative} calibration of the signal level coming out from the quantum amplifier.
In 2024, the project {\it Quantum Technologies for Neutrino Mass} developed a sensing apparatus to measure the neutrino mass with a sensitivity of 10 $meV/c^2$ \cite{amad2024determiningabsoluteneutrinomass}. Improving this sensitivity is here a key ingredient and relies on quantum amplifiers.
As well, one candidate for Dark Matter is the {\it axion}, 
which is today actively sought 
in cryogenic microwave setups employing TWPA read-out \cite{Axions}.

Among quantum sensing, 
nano-mechanics, and specifically opto-mechanics, is an extremely active field of research tackling fundamental questions at the boundary between quantum mechanics and general relativity. Proposals for direct {\it graviton} detection \cite{PikovskiNature}, or signatures of gravitational effects based on microwave electro-opto-mechanics \cite{MikaSpheres,SteeleDecoh,lairdAVS,MikaNear} are today under consideration.
For these, absolute calibration gives access quantitatively to motion amplitude and mechanical mode population.
In the first place it leads to 
mode temperature definition, but more importantly it is mandatory for a 
proper extraction of parameters quantifying {\it entanglement} \cite{MikaEntangle}, or more speculatively the "effective heating" appearing in collapse models \cite{DiosiPRL2015}.  \\

In the present article we report on the quantitative characterization of a TWPA-based near 
quantum-limited detection setup.
This is performed 
on a unique microwave optomechanical platform, able to cool down the system to the coldest possible temperatures achievable for condensed matter (by means of nuclear demagnetization \cite{XinPRAppl}).
The opto-mechanical measurements are performed using a drum-shaped aluminum membrane coupled to a superconducting cavity, 
which reached mode temperatures as low as 500$~\mu$K  \cite{dylan} using the same setup, before the installation of the TWPA. We do not intend here to explore ground state properties, and will remain at temperatures above 1$~$mK.
The TWPA properties are analyzed on the basis of the Two Level Systems (TLS) theory \cite{tls,tls2}, phenomenologically reproducing results known for superconducting cavities \cite{Capelle}.  
It turns out that TLSs seem to play an important role in the dynamics of TWPAs, as recently shown with {\it echo experiments} \cite{benji}.
We demonstrate that we can extract the population of the mechanical mode within $\pm 20~\%$ by means of an {\it absolute} calibration, reaching a population of 5$~$phonons at 4$~$mK.

\section{\label{sec:level2}Experimental setup}

\begin{figure}[h]
        \hspace*{-0.7cm}
        \includegraphics[width=0.55\textwidth]{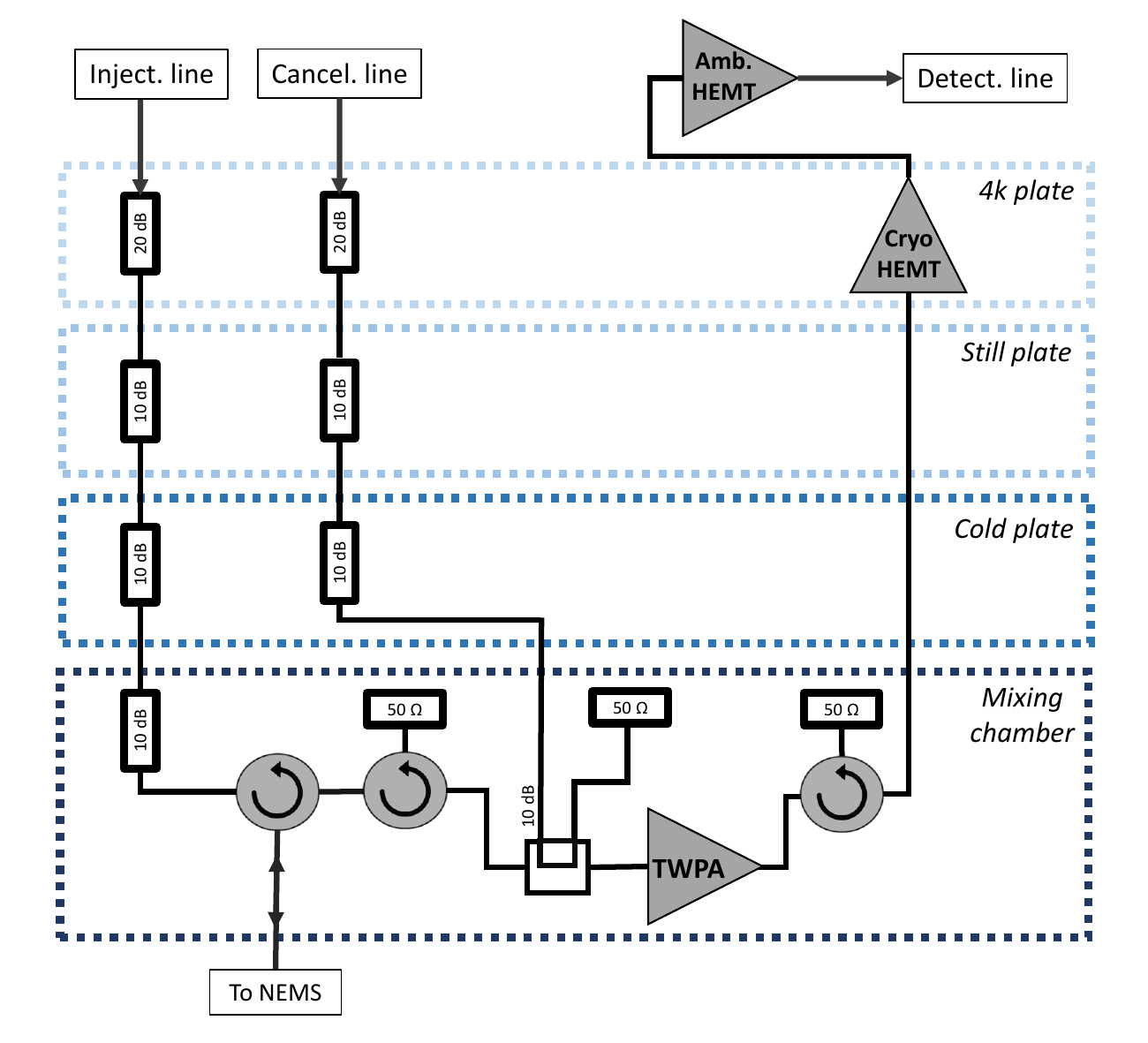}
    \caption{Low temperature microwave circuitry: after splitting the optomechanical pump signal into injection line and cancellation line, 
    both signals are sent into the cryostat through attenuators and circulators. Both are recombined before the TWPA to avoid any saturation. Another circulator is implemented downstream to protect the TWPA from 
    the noise generated by the cryogenic HEMT.
    The total attenuation in injection and gain in detection are calibrated within an uncertainty of about $\pm1$ dB. The different anchoring stages of the dilution unit are specified. }
    \label{fig:1}
\end{figure}

 The %
 microwave 
 experimental platform is 
installed on the dilution unit of a nuclear demagnetization cryostat \cite{XinPRAppl}. 
The schematics is reproduced in Fig. \ref{fig:1}; it is the same setup as the one of Ref. \cite{Andrew}.
  A pump signal (which is required for the optomechanical schemes discussed in the next Section) 
  is sent into the cryostat through attenuators that eliminate the thermal 
  noise.
Whenever an extra small probe tone is required for the experiment, it is injected through the same injection line. 
  A double circulator is added in order to protect the sample cell from any signals traveling backwards from the amplifiers 
  (noise, or reflected waves due to mismatched components). 
  A second signal, created by splitting the pump tone at room temperature and amplitude-and-phase controlled, is injected in a separate line towards the amplifiers in order to suppress the strong pump. The two signals are combined by 
 a directional coupler. 
  Efficient pump cancellation 
  is especially important here, since the TWPA saturates at extremely low input powers, typically about  $-110~$dBm \cite{PhysRevApplied.11.034014}.
 
The detection is performed by means of a Zurich Instruments\textsuperscript{\textregistered} 
 lock-in amplifier. The 
room temperature HEMT 
output signal is down-converted with a mixer and a  
 Local Oscillator (LO) \cite{XinPRAppl,Andrew}.
  The detection line that brings the signal to the lock-in is composed of a room temparature HEMT (a conventional Low Noise Factory\textsuperscript{\textregistered} device), and a last generation Low Noise Factory\textsuperscript{\textregistered} cryogenic HEMT with $T_{noise} \approx 2~$K.
The first stage of amplification is performed with an  
"Argo" Silent Waves\textsuperscript{\textregistered} TWPA, with an insertion loss of  about $3 -4 ~$dB around $5~$GHz.
This component enables us to bring the detection limit as close as possible to the Standard Quantum Limit (SQL). 
It requires a pump tone to function, which is sent through the cancellation line (Fig. \ref{fig:1}). It also must be protected against the noise of the HEMT, which is why an extra circulator is present on the schematics between the two cryogenic amplifiers.


The NEMS device used in this study, which is mounted on the copper nuclear demagnetization stage, is the one already described in Ref. \cite{dylan}. 
It is made of a 2-dimensional aluminum drum-shaped structure embedded in an aluminum microwave cavity. The  cavity, resonating at $\omega_c = 2\pi\cdot 5.154~$GHz, is reflectively coupled with an external damping rate $\kappa_{ext} = 2\pi \cdot 180~$kHz (temperature and power independent), 
leading to a  (best) total damping rate $\kappa_{tot} = 2\pi \cdot 450~$kHz (temperature {\it and} power dependent, see following Sections). 
The fundamental flexural mode of our NEMS resonates at $\Omega_m = 2\pi\cdot 15.13~$MHz, with a 
damping rate $\Gamma_{m} = 2\pi \cdot 420~$Hz at the lowest temperatures.
The optomechanical coupling between NEMS and cavity 
is $g_0 = 2\pi \cdot 220~$Hz.


\section{Basics of optomechanics}

The microwave setup discussed in the present article is meant to be used for optomechanical experiments.
Optomechanics is based on the interaction between light, confined in a cavity, and a moving element (a mirror): the so-called {\it radiation pressure} \cite{Aspel}. 
We deal here with the microwave version of it, where the mirror is replaced by a moving capacitor \cite{Regal2008}.

\begin{figure}[h]
    \vspace*{-0.2cm}
    \hspace*{-0.5cm}
    \includegraphics[width=0.52\textwidth]{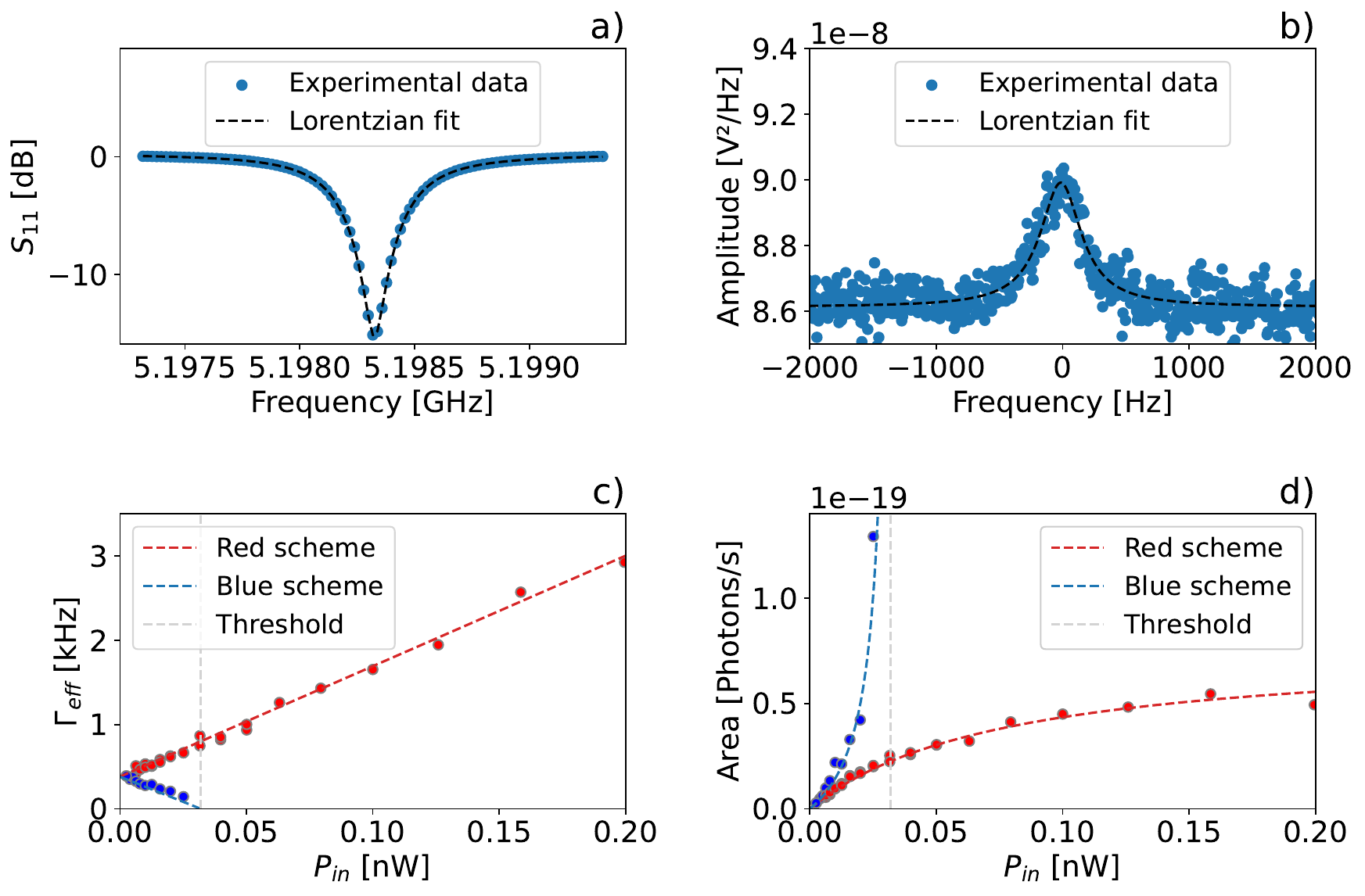}
    \caption{Typical 
    optomechanical data taken at $T = 20~$mK.
    a) Cavity measurement using a large probe tone ($1 \cdot 10^{-4}~$nW, and no pump); the resonance is fit with a Lorentzian function defined in Eq. (\ref{S11}); note the logarithmic scale. 
    b) Raw mechanical peak obtained with the blue pumping scheme ($P_{in} = 2.5 \cdot 10^{-3}~$nW), and its Lorentz fit. 
    Integrating this peak leads to the sideband area $A_{sdb}$ expressed in photons/s, and described by Eq. (\ref{sidebandA}), taking into account the proper gains and conversion factors.
    c) $\Gamma_{e\!f\!f}$ measurement using blue and red schemes (respectively anti-damping/damping processes)
    as a function of the on-chip applied power (corrected from injection attenuation). The absolute value of the slope directly gives $g_0$, the optomechanical coupling constant, through Eq. (\ref{g0}).
   d) Measured signal area 
    for blue- and red- pumping configurations, with fits corresponding to Eq. (\ref{sidebandA}). 
       In c) and d) the dashed vertical corresponds to the threshold for self-oscillation (see text). }  \label{fig:2}
\end{figure}

Two conventional 
single-tone schemes are used in this study, depending on the pump frequency $\omega_p$: the "blue" ($\omega_p = \omega_c + \Omega_m$) and the "red" ($\omega_p = \omega_c - \Omega_m$) pumping configurations, 
in analogy with Raman scattering processes.
The motion of the 
capacitor imprints sidebands in the 
microwave %
optical signal, which can be detected in the output field.
The back-action of the cavity on the NEMS leads in the "blue" case to an amplification of the motion, and anti-damping, while in the "red" case we obtain cooling, and extra damping. 
These are all the more prominent as the pump power $P_{in}$ is increased.
For our particular NEMS, blue and red pumping are demonstrated in Fig. \ref{fig:2}, together with a microwave cavity and a NEMS sideband peak measurement. \\


The reflexion measurement $S_{11}$ demonstrating the cavity dip is fit using {\it its magnitude in dB},  Eq. (\ref{S11}) \cite{Ilyasingle}:
\begin{eqnarray}
  \!\!\!\!\!\!\!  S_{11,{\mbox dB}} &=& 20 \log \Bigg|\frac{\dfrac{Q_{ext} 
    - 2Q_{tot}}{Q_{ext} 
    }{+ 2i\,Q_{tot} \left(\dfrac{\omega - \omega_c}{\omega_c}\right)}}{1 + 2i\,Q_{tot} \left(\dfrac{\omega - \omega_c}{\omega_c}\right)}
    {} \Bigg| , \label{S11} \\
    Q_{tot} &=& \Bigg(\frac{1}{Q_{ext}} + \frac{1}{Q_{in}}\Bigg)^{-1} , \label{eq:1}
\end{eqnarray} 
in which we defined the external  quality factor $Q_{ext}=\omega_c/\kappa_{ext}$, the total  quality factor $Q_{tot}=\omega_c/\kappa_{tot} $, and the internal quality factor $Q_{in}=\omega_c/\kappa_{in} $, 
having $\kappa_{tot}  = \kappa_{ext} + \kappa_{in} $.
It thus
allows us to extract independently $\omega_c,~\kappa_{tot}$ and $\kappa_{in}$ with good accuracy.
The $\kappa_{in}(T,P_{in})$ rate corresponds to all relaxation processes occurring within the cavity, and it does depend on both $T$ and $P_{in}$: we carefully characterize it in the next Section.
Note however that Eq. (\ref{S11}) leads to two possible solutions for the $\kappa_{ext}, \kappa_{in}$ couple, at fixed $\kappa_{tot}$: either overcoupled $\kappa_{ext}>\kappa_{in}$, or undercoupled $\kappa_{ext}<\kappa_{in}$. This ambiguity is {\it lifted} as soon as we consider the full quantitative modeling of the setup, as discussed below. \\

The mechanical damping rate $\Gamma_{e\!f\!f}$ (linewidth of the Lorentz peak in Fig. \ref{fig:2} b) is fit to:
\begin{eqnarray}
\!\!\!\!\!\!\!\!\!\!\!\!\!\!    \Gamma_{e\!f\!f}(T,P_{in}) &=& \Gamma_{m}(T) \pm \Gamma_{opt}(T,P_{in})   ,\label{eq:1}  \\
\!\!\!\!\!\!\!\!\!\!\!\!\!    \Gamma_{opt}(T,P_{in}) &=& \frac{4g_0^2}{\kappa_{tot}(T,P_{in})}   \frac{\kappa_{ext}\,P_{in}}{\hbar \,\omega_c\bigg[\Omega_m^2 + \! \left(\frac{\kappa_{tot}(T,P_{in})}{2}\right)^{\!2} \,\!\! \bigg]} \!, \label{g0} 
\end{eqnarray}
with the sign $\pm$ standing for the two pump schemes %
(and $\hbar$ is Planck's reduced constant). 
For blue pumping ($-$ sign), the damping goes down until it reaches zero: at that point, the mechanical mode starts to self-oscillate \cite{Del2024,DylanPRR}.
The area of the mechanical sideband peak $A_{sdb}$ (in photons/s on chip) is given by \cite{XinPRAppl}:
\begin{eqnarray}
    A_{sdb}(T,P_{in})&=& 
    A_{0}(T)\,G_{opt}(T,P_{in})  , \label{sidebandA}\\
    G_{opt}(T,P_{in}) &=& \frac{\Gamma_m (T)}{\Gamma_{e\!f\!f}(T,P_{in})} , \label{amplifdeplif}
\end{eqnarray} 
with $G_{opt}$ the optomechanical gain function, and $A_{0}$
the original peak area in the absence of amplification/deamplification. 
This quantity can be converted from photons/seconds into an area 
 $A_{ph}$ expressed in {\it phonons}, which corresponds to the occupation number of the mechanical mode:
\begin{eqnarray}
        A_{0}(T,P_{in}) &=& A_{ph}(T)\,\mathcal{M}(T,P_{in})\,P_{in} , \label{eq:1bis} \\
        \mathcal{M}(T,P_{in}) &=& \frac{4\,g_0^2\,\kappa_{ext}^2}{\hbar \,\Omega_m^2\,\omega_c \, \kappa_{tot}^2(T,P_{in})} . \label{eq:2M}
\end{eqnarray} 
This 
area $A_{ph}$ is a crucial parameter: at  temperatures $T \gg  \hbar \Omega_m/k_B$ ($k_B$ Boltzmann's constant), 
it is simply $\propto T$ and proves that the mode is properly thermalized. In the reverse limit, it {\it displays sideband asymmetry} \cite{dylan}: the area obtained with the red scheme goes to zero, while the blue one saturates at exactly one phonon. This is a direct signature of quantum mechanics. 
Quantitative fits therefore rely on the knowledge of $\kappa_{tot}$ which appears in all Eqs. (\ref{g0},\ref{amplifdeplif},\ref{eq:2M}). The conversion factor $\mathcal{M}$ depends directly on $\kappa_{ext}$, so that a proper calibration defines it (overcoupled or undercoupled) without ambiguity.

\section{Cavity properties}
\label{cavcap}

\begin{figure}[h]
    \centering
    \includegraphics[width=0.46\textwidth]{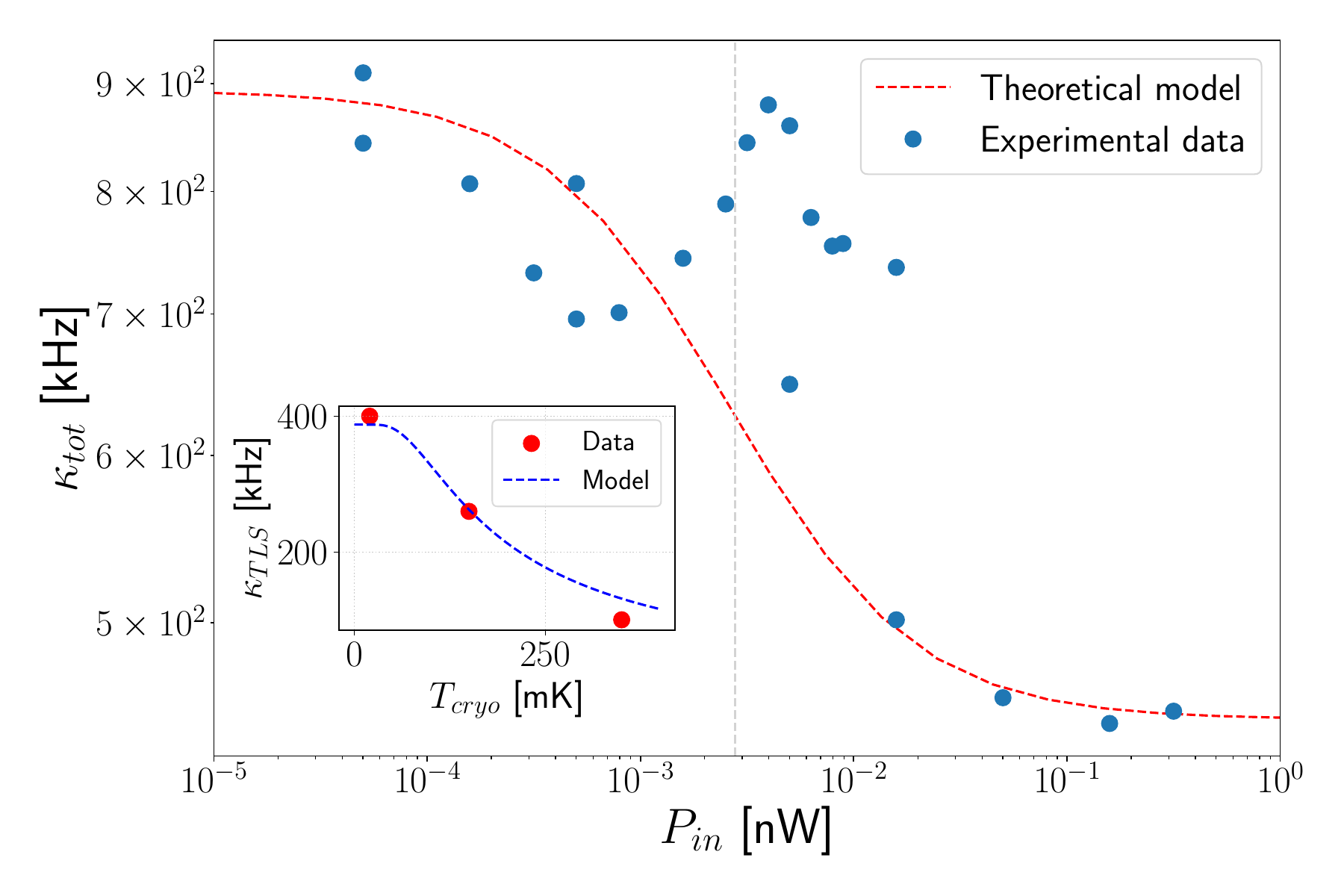}
    \caption{Main: Total damping rate of the cavity measured as a function of pump applied power (blue scheme, using a small probe tone, $5\cdot10^{-7}~$nW) at $T = 150$ mK. 
    The fit corresponds to 
    Eq. (\ref{kappatot}). The gray vertical corresponds to the inflection point defined by the critical power $P_{cav}^0$. A dramatic drop ($\sim 50$ \%) in the total damping rate is observed while increasing the power (see text).
    Inset: temperature dependence demonstrating the characteristic $\tanh$ Eq. (\ref{tanh}) shape (see text). }
    \label{fig:3}
\end{figure}

We illustrate in Fig. \ref{fig:3} the microwave cavity total damping rate $\kappa_{tot}$ as a function of pump power $P_{in}$ (main) and $T$ (inset). 
The key features of this behavior are 
understood by means of the theoretical model of Ref. \cite{Capelle}, involving the saturation of TLSs in 
the constitutive material of the superconducting circuitry.

The TLSs are coupled to the microwave field through their electric dipolar moment. They can absorb energy, and get excited. At high powers (in steady-state drive) or high temperatures, they reach their thermal equilibrium with 1/2 population in both excited and ground states: they are then unable to absorb energy anymore, and their contribution to the cavity relaxation channels disappears.
We can fit these power and temperature dependencies 
with only two fitting parameters: $P_{cav}^0(T)$   the temperature-dependent critical power required to saturate the TLSs, 
and $\kappa_{TLS}^0$ the overall TLS contribution: 
\begin{eqnarray}
\!\!\!\!\!\!\!\!\!\!\!\!\!\!\!\!\! \kappa_{in}(T,P_{in}) &\!\!\,=\!\!& \kappa_{TLS}(T)\frac{P_{cav}^0(T)/P_{in}}{1+P_{cav}^0(T)/P_{in}} + \kappa_{BCS}(T) , \label{kappatot}  \\
\!\!\!\kappa_{TLS} (T) &\,\,\!=\!\,&  \kappa_{TLS}^0  \tanh \left(\frac{\hbar \omega_c}{2k_B T}\right) . \label{tanh}
\end{eqnarray} 
Measurements are performed with a small probe tone sweeping the cavity, while increasing the power of a much stronger blue detuned pump tone.
Fits are presented in Fig. \ref{fig:3} (the {\it detuned signal} expressions from Ref. \cite{Capelle}).
The temperature dependence of $P_{cav}^0(T)$, which corresponds to the ability of TLSs to absorb energy, is related to their characteristic rates $\Gamma_1(T)$ (relaxation) and $\Gamma_2(T)$ (dephasing)
 \cite{Capelle}. We shall not comment on this any further, which is outside of our scope. 
The relevant point here is that the TLS model  gives a good account of the measured $\kappa_{tot}$. \\
In Eq. (\ref{kappatot}), another term appears: $\kappa_{BCS}(T)$. It actually refers to the intrinsic superconducting cavity damping rate, which grows as we increase temperature. 
This is captured by the following equation:
\begin{equation}
    \kappa_{BCS}(T) = \kappa_{dielec}^{0} +\alpha \frac{T_c}{T} \exp^{-\frac{\Delta(0)}{k_B T}} ,     \label{eq:supra}
\end{equation}  
in which the first constant $\kappa_{dielec}^{0}$ corresponds to the dielectric losses arising from the underlying substrate. The temperature-dependent term is the Mattis-Bardeen expression \cite{Bardeen}, valid typically for $T < T_c/2$, and corresponds to the growth in free electrons density (broken Cooper pairs) as the temperature is increased.
$\Delta(0) \approx 3.3 \, k_B T_c$ with $T_c$ the aluminum film critical temperature, and $\alpha$ is a fit parameter (which is related to aluminum properties and the cavity geometry). \\

\begin{figure}[h!]
    \centering
    \includegraphics[width=0.48\textwidth]{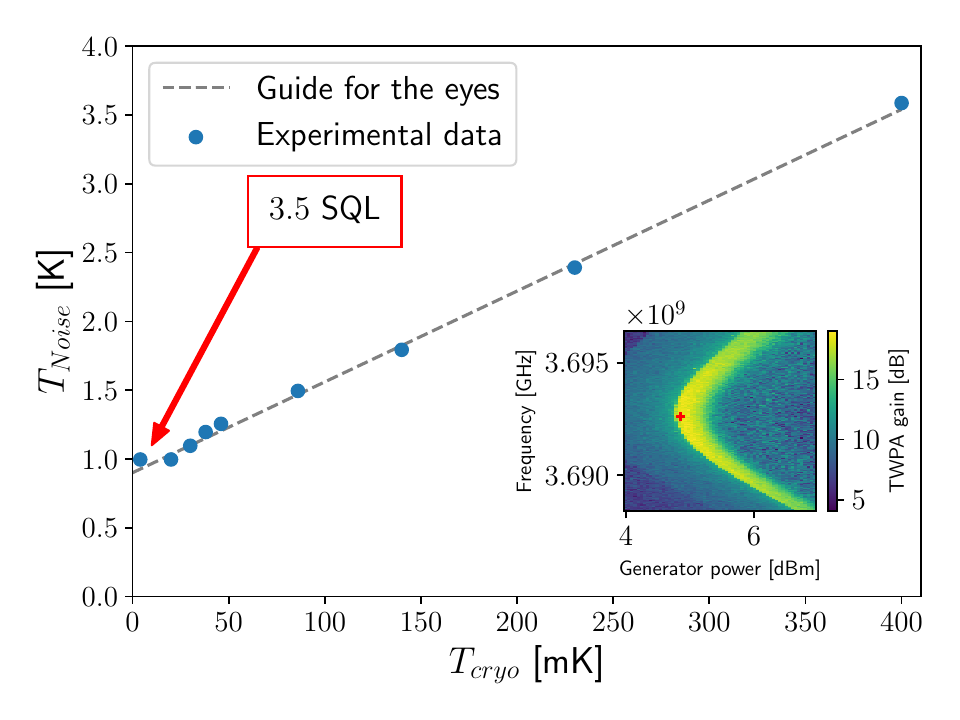}
    \caption{Main: background noise measurements as a function of cryogenic temperature. The $T_{cryo} \rightarrow 0~$K interpolation corresponds to the fundamental limit of the detection  system, here about $3.5~$SQL. 
    Note that the dashed line is only a guide: the temperature-dependence being due to both the last 50$~\Omega$
 load of the setup, but also to the distributed losses within the TWPA itself, which we do not intend to model here.
    Inset: Typical result of a TWPA pump parameter tuning scan at $T = 20$ mK. The yellow part of the graph corresponds to the highest values of TWPA gain, while the red cross stands for the optimal set of parameters chosen (see text for further details), giving a gain of $18 \pm 1~$dB.}
    \label{fig:4}
\end{figure}

Below about 300$~$mK, the $\kappa_{BCS}$ term is negligible.
At our highest temperatures (about $400~$mK), the total damping rate $\kappa_{tot}$ reaches $\sim 3~$MHz. 
At the lowest temperatures, the TLS saturation  
can lead to about $50~\%$ change in cavity damping, 
which is significant 
and obviously needs to be considered for our quantitative approach. 
Besides, we show in the next Section that {\it the same phenomenon} happens in the TWPA: and a phenomenological fitting adapted from Ref. \cite{Capelle} describes well the experimental results.

\section{Twpa properties}
The TWPA requires a pump tone with an adequate pump power $P_{TW\!P\!A}$ and pump frequency $\omega_{TW\!P\!A}$ to function properly.
The figures of merit 
are the TWPA gain, and also the TWPA added noise, which both depend on the working point. 
As importantly, we require the found settings to be {\it as stable as possible}, in order to enable long measurement times without any fluctuations.
We therefore proceed by mapping out the 
parameter space of the TWPA pump generator. 
A typical such measurement is given in Fig. \ref{fig:4} (inset). 
We slowly scan a small probe tone, whose transmission is measured with the TWPA pump on and off. 
The ratio defines the plotted gain.
The optimal set of parameters is defined by considering the zone with maximal TWPA gain, 
and also the lowest local variation: this is represented by the red cross in Fig. \ref{fig:4} (inset). 
The mechanisms behind instabilities are not fully understood \cite{Planat}; so in our case, for each change of cryostat settings (fridge base temperature, or main demagnetization magnet field) we redo the procedure.

In Fig. \ref{fig:4} (main graph) we thus also plot the amplification chain total noise, referenced at the input of the TWPA. 
The graph is obtained from a careful calibration of the losses of all components inserted in the circuit, and the gains of the two HEMTs. The extracted noise figure had been validated against the {\it 
cold/hot load technique}: 
the noise measured from a 50$~\Omega$ load bolted at the 4$~$K stage is compared to the one of a similar load located at the 10$~$mK stage. For details on the procedure, see Ref. \cite{IlyaMeso}.
We reach at the lowest temperatures about 3.5$~$ photons system
 noise at 5$~$GHz, which is about the best figure for this amplifier.
 
The on/off TWPA gain obtained is $G = 18 \pm 1$ dB, while the signal-to-noise improvement (taking into account the insertion loss) is about a factor 10, as compared to the HEMT alone. As the temperature of the TWPA stage rises, the noise increases gradually while the gain suddenly drops above typically 300$~$mK; it is actually of order 1 at the highest temperatures. \\

It turns out that the same phenomenon of TLS energy absorption, which we have seen in the previous Section for a {\it single} microwave cavity, is also present in the TWPA
which is constituted of {\it a chain of resonators} \cite{Planat2}. 
And this happens already when the TWPA pump is off: it appears as an extra {\it power-dependent insertion loss}.

\begin{figure}[h!]
\hspace{-0.5cm}
    \includegraphics[width=0.50\textwidth]{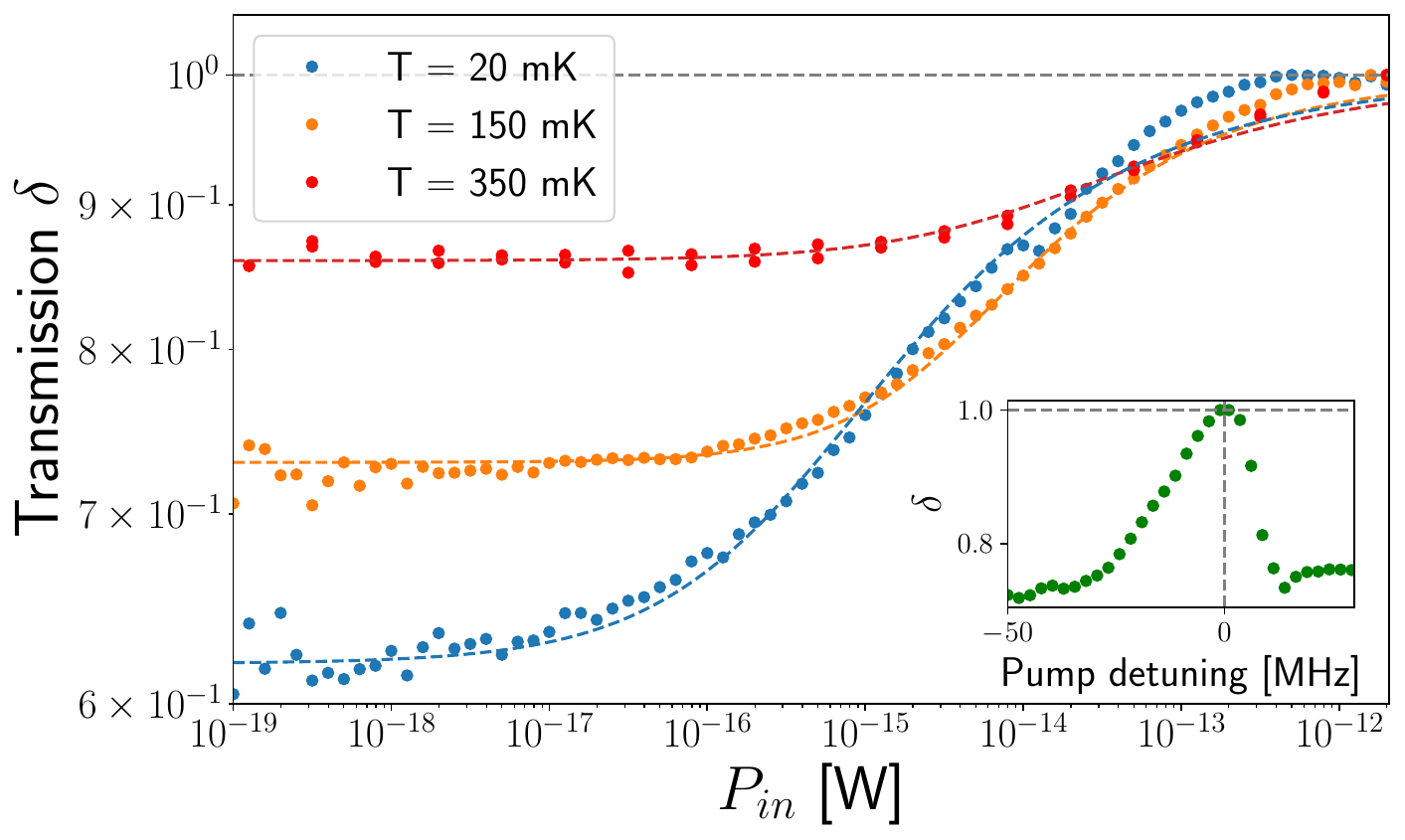}
    \caption{Main: TWPA transmission 
    around 5$~$GHz (with TWPA pump off)  measured as a function of applied probe power, for various temperatures (and optomechanics pump off). 
     Fitting functions correspond to Eqs.  
     (\ref{eq:fitTWPA},\ref{eq:fitTWPAT}). The drop reaches about $40~\%$ at $20~$mK for low powers (see text). 
    Inset: TWPA transmission  at 150$~$mK (with TWPA pump off) measured with constant small probe power (about $ 3\cdot10^{-17}~$W), but as a function of the detuning of a large pump tone from the probe (optomechanics pump, about $2\cdot10^{-12}~$W). While detuning the pump, the losses reappear (for details, see text). 
    }
    \label{fig:5}
\end{figure}

This is illustrated in Fig. \ref{fig:5}. The TWPA pump is kept off, and we measure the transmission through the full setup with a small probe tone, slightly detuned from the microwave cavity in order not to mix up the imprint of TLSs arising from the TWPA's and the optomechanics' chip. 
The specific features of the TLS saturation are observed: at low probe power, the transmission $\delta(T,P_{in})$ degrades because of energy absorption by the TLSs, see  Fig. \ref{fig:5} main plot. 
This can be reproduced by adapting phenomenologically 
the expressions given in Ref. \cite{Capelle} for a {\it tuned injected signal}, 
introducing fit parameters $\lambda_0$, $\beta$:
\begin{eqnarray}
    \delta(T,P_{in}) &=& 1-\frac{\lambda(T)}{\sqrt{1+\left(\dfrac{P_{in}}{P_{twpa}^0(T)}\right)^{\!\beta}}} 
   ,  \label{eq:fitTWPA} \\
   \lambda(T) &= & \lambda_0 \,\tanh\left(\dfrac{\hbar \omega_c}{2k_B T}\right) , \label{eq:fitTWPAT}
\end{eqnarray} 
see the dashed lines in Fig. \ref{fig:5}. 
$\beta \approx 1$ accounts for a slightly inhomogeneous power distribution \cite{tls,tls2}.
As well, keeping a small probe for the measurement, and using the optomechanics pump to saturate TLSs (with a large enough power, 
and the {\it cancellation signal} off), we study the frequency detuning dependence in Fig. \ref{fig:5} (inset).
When the pump is close to the probe, we recover the full transmission. However, while shifting away the pump from the probe, we see that the transmission degrades again: the saturation becomes inefficient outside of a bandwidth (here, about 25$~$MHz) characteristic of the TLSs relaxation/dephasing rates \cite{Capelle}. We do not intend to propose a phenomenological fit for this, because of its peculiar asymmetric shape which we cannot reproduce.
A proper modeling (that could also account for the parameters $\lambda_0, \beta$), would have to take into account the distributed nature of the TWPA structure; this is a rather difficult task \cite{PhysRevApplied.12.034054}.
Furthermore, $P_{twpa}^0$ plays the role of the $P_{cav}^0$ discussed in the previous Section, and as such displays a temperature dependence. 

\begin{figure}[h!]
    \hspace*{-0.6cm}
    \includegraphics[width=0.50\textwidth]{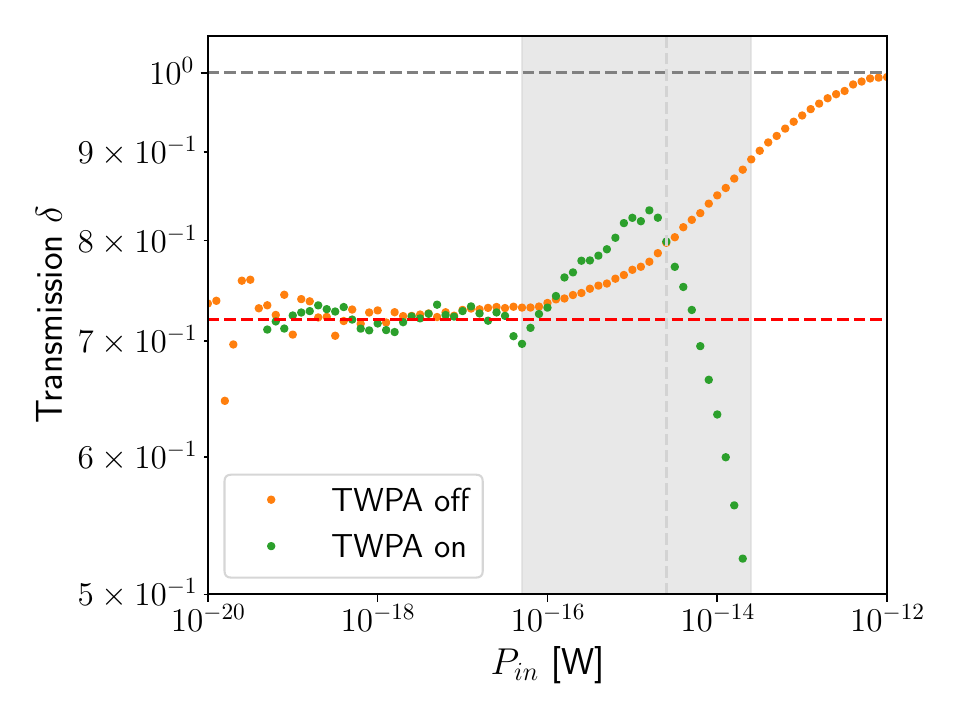}
    \caption{Comparison of TWPA transmission versus power with/without TWPA pump at $T = 150~$mK. Orange points correspond to Fig. \ref{fig:5} (TWPA off), while green points are measured turning on the TWPA pump (and scaling properly for the TWPA gain). 
   The vertical band illustrates the saturation zone of the TWPA (see text for details).}
    \label{fig:6}
\end{figure}

Finally, Fig. \ref{fig:6} compares the transmission while turning on/off the TWPA pump.
%
The two datasets start at the same level, by definition (this is actually how the gain is defined). But as the probe power is increased, we see a clear difference: when the TWPA pump is on, the signal very rapidly degrades.
This is marked by the shaded region in Fig. \ref{fig:6}: a first kink is visible around $10^{-16}~$W, and then the signal drops dramatically above $10^{-14}~$W. 
This is nothing but the TWPA amplifier's saturation \cite{PhysRevApplied.11.034014,arxivPlus}. What Fig. \ref{fig:6} tells us is that it is {\it impossible} to saturate the TLSs while keeping the TWPA gain optimum: the TWPA saturates first.
As a result, if one wants a quantitative measurement %
 of the transmitted signal amplitude, one {\it has to take} the TWPA TLSs 
 into account: there is no way around here. To do so, phenomenological fits such as Eqs. (\ref{eq:fitTWPA},\ref{eq:fitTWPAT}) are enough, and will be applied in the next Section in the framework of optomechanics.


\section{Calibrated optomechanics signals}

\begin{figure}[h]
    \centering
    \includegraphics[width=0.48\textwidth]{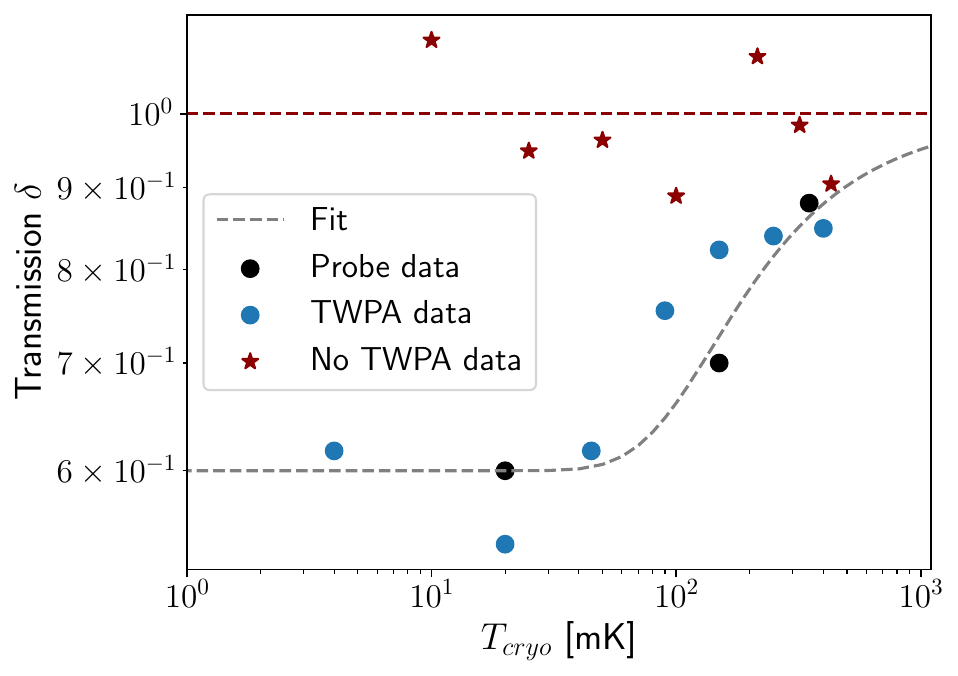}
    \caption{TWPA transmission as a function of cryogenic temperature. "Probe data" correspond to 
    points measured with a small probe tone 
    (and TWPA pump off), see Fig. \ref{fig:5} and previous Section.
    The other datasets are obtained from optomechanical data, plotting $A_{ph}/n_{ph}$.  
    The "No TWPA data" corresponds to a set taken on the same microwave platform, without the TWPA (and with another HEMT) \cite{dylan}. The rest of the wiring was indentical.
    "TWPA data" corresponds to this run, with TWPA pump on. The gray dashed line is our fitting function, Eqs. (\ref{eq:fitTWPA},\ref{eq:fitTWPAT}), see text.}
    \label{fig:7}
\end{figure}

We now turn to the use of this microwave platform for optomechanics measurements.
From Eq. (\ref{eq:1bis}), we extract the area $A_{ph}$. 
This is properly performed thanks to the cavity TLS analysis of Section \ref{cavcap}, for various input powers and temperatures. 
At 
temperatures $T > 1~$mK, blue and red pumping schemes are essentially equivalent and 
the value should match the Bose-Einstein phononic population:
\begin{equation}
n_{ph}(T) = \frac{1}{\exp[\hbar \Omega_m/(k_B T)]-1} , \label{bose}
\end{equation}
which reduces to $k_B T/(\hbar \, \Omega_m)$ for $n_{ph} \gg 1$. We therefore plot in Fig. \ref{fig:7} the ratio $A_{ph}/n_{ph}$
 (measured phonon population normalized to expectation) as a function of cryostat temperature $ T_{cryo}$. 
Note that the signal power is much smaller than $10^{-16}~$W, and thus cannot saturate neither the TWPA, nor the TLSs (See Fig. \ref{fig:6}). So without any correction, this should simply reflect the temperature dependence of the TWPA transmission $\delta$, Eqs. (\ref{eq:fitTWPA},\ref{eq:fitTWPAT}). To prove the point, we compare this to the same ratio obtained in a run with no TWPA, on the same cryostat \cite{dylan}: 
in this case, $A_{ph}/n_{ph}$ remains around 1 at all temperatures. The wiring was exactly the same, apart from the HEMT which was an old Caltech\textsuperscript{\textregistered} model with much higher noise level (but essentially the same gain). We also plot the probe data of the previous Section. 

The data acquired with the TWPA pump on clearly follow our proposed fit expression in Fig. \ref{fig:7}, 
in agreement with the probe data (gray dashed line). 
A signal loss %
attributed to TLSs  
of up to 40$~\%$ affects the measurements, and needs to be taken into account for quantitative fits. 
Let us comment on the error bars of this final graph. 
The temperature measurement $T_{cryo}$ is performed at $T>10~$mK by means of calibrated carbon resistors, and of a Magnicon MFFT noise thermometer, regulating the temperature of the cryostat's mixing chamber. We estimate the error bar there smaller than $\pm5~\%$. For the lowest temperature point at $4~$mK, achieved by nuclear demagnetization, the error is much larger, about $\pm20~\%$ because these thermometers start to saturate on our setup. We rely on the magnetic temperature of the nuclear spins, which is estimated here from repeated upmagnetization/demagnetization cycles.   
We should also point out that the conversion factor $\mathcal{M}$, Eq. (\ref{eq:2M}), depends strongly on parameters which must be measured, mainly $g_0$, $\kappa_{ext}$ and $\kappa_{tot}$ which at best are known within $\pm5~\%$. This generates an {\it absolute} error of about $\pm 30~\%$. Since the data in Fig. \ref{fig:7} does fall at high temperature on the right normalization value (dashed horizontal at 1), we conclude that the overall estimate of $\mathcal{M}$ is particularly good. \\

Finally, it is clear on Fig. \ref{fig:7} that the data reproducibility is about $\pm20~\%$. This is actually {\it intrinsic} to the mechanical device, and arises from $1/f$-type fluctuations that limit the quality of the data averaging \cite{dylan}.
Such noises, which ultimately limit our {\it relative} error, deserve further studies.
Relying on the work presented here, we can calibrate the setup over two decades in temperature, for any microwave power (and optomechanical scheme, %
as defined from the pump detuning). 
We end up at $4~$mK with an inferred population of about {\it 5 phonons}, reproducibly measured within $\pm20~\%$, consistent with Eq. (\ref{bose}).



\section{Conclusions}

%

We report on the thorough calibration of a microwave optomechanical quantum-limited setup, constructed around a state-of-art Traveling Wave Parametric Amplifier (TWPA).
The intrinsic losses due to Two-Level-Systems (TLSs) present in the optomechanics cavity are fit to the theory of Ref. \cite{Capelle}.
We demonstrate that a similar feature exists in the TWPA, which we fit by adapting phenomenologically these expressions.

This TLS dissipative channel appears as an extra {\it insertion loss} for the TWPA, which is {\it power and temperature dependent}. It is fit in order to produce a quantitative calibration of the detection line, {\it in the full parameter space}: 
one just needs to normalize the measured signal (for us, the optomechanical sideband area) with Eqs. (\ref{eq:fitTWPA},\ref{eq:fitTWPAT}).
We demonstrate that the TWPA pump {\it does not} saturate the TLSs (the TWPA itself saturates first, and becomes unusable): the normalization procedure is therefore the only way to take the TLS effect into account, 
which can lead to a loss in signal of up to a factor 2 here.
As such, we can extract from the measurements the mean phonon population 
of our nano-mechanical device from $4~$mK to $400~$mK, with an achieved uncertainty of $\pm20~\%$ over the full temperature range. 
This scatter is actually neither related to the TWPA nor the method we employ, it is actually internal to the optomechanical device  \cite{dylan}.

We believe that our TWPA calibration method is particularly useful for quantum sensing experiments, especially in the field of microwave optomechanics. 
Future experiments near the ground state of motion, for which a quantum-limited amplifier is mandatory, will benefit from our results. The careful study of thermodynamics properties at ultra-low temperatures, and especially {\it{fluctuations}} of mechanical properties should shed a new light on quantum nano-mechanics \cite{IlyaMeso,EddyAVSQ}.
A proper ab-initio calibration guarantees quantitative measurements, in particular ensuring the compatibility between low and high temperature data. 
It eventually relies {\it only} on microwave properties, not on a specific normalization point: in other words, mechanical motion (or phonon population) is derived in absolute units, not as a relative comparison to a fixed temperature reference. 



\vspace*{0.2cm}
\section*{Acknoledgments}

We acknowledge the help of the N\'eel Cryogenics team, and funding from the ANR Grant MORETOME No. ANR-22-CE24-0020-01. 
This project has received funding from European Union’s Horizon Europe 2021-2027 project TruePA (grant agree-
ment number 101080152), and from the French ANR-22-PETQ-0003 grant under the ’France 2030’ plan.
We also received funding from the 
Leverhulme Trust 
under Research Project Grant Ultra-Cool Mechanics (RPG-2023-177).
%
The work was performed as part of the Academy of Finland Centre of Excellence program (project 336810). 
We acknowledge funding from the European Union’s Horizon 2020 research and innovation program under the QuantERA II Programme (13352189).
The research leading to these results has been conducted in the framework of the European Union's Horizon 2020 Research and Innovation Programme, under grant agreement No. 824109, the European Microkelvin Platform (EMP).


\end{document}